%% file: main.tex
\pdfoutput=1
 \documentclass[aps, prd, twocolumn, nofootinbib, longbibliography, superscriptaddress]{revtex4-2}

\newcommand{\aap}{Astronomy and Astrophysics}
\newcommand{\mnras}{Monthly Notices of the Royal Astronomical Society}
\newcommand{\jcap}{Journal of Cosmology and Astroparticle Physics}
\newcommand{\apjl}{The Astrophysical Journal}	
\newcommand{\apjs}{The Astrophysical Journal Supplement Series}

\usepackage{natbib}
\usepackage{amsmath}
\usepackage{float}
\usepackage{hyperref}
\hypersetup{colorlinks=true, allcolors=blue}
\usepackage{graphicx}
\usepackage[english]{babel}

\newcommand{\be}{\begin{equation}}
\newcommand{\ee}{\end{equation}}
\newcommand{\ba}{\begin{eqnarray}}
\newcommand{\ea}{\end{eqnarray}}

\newcommand{\LCDM}{$\Lambda$CDM}

\begin{document}

\title{A Python compressed low-$\ell$ \textit{Planck} likelihood for temperature and polarization}

\author{Heather Prince}
\email{heatherp@princeton.edu}
\affiliation{Department of Astrophysical Sciences, Peyton Hall,
Princeton University, Princeton, New Jersey 08544, USA }

\author{Jo Dunkley}
\affiliation{Department of Astrophysical Sciences, Peyton Hall,
Princeton University, Princeton, New Jersey 08544, USA }
\affiliation{Joseph Henry Laboratories of Physics, Jadwin Hall, Princeton University, Princeton, NJ, USA 08544}

\begin{abstract}
We present \texttt{Planck-low-py}, a binned low-$\ell$ temperature and $E$-mode polarization likelihood, as an option to facilitate
ease of use of the \textit{Planck} 2018 large-scale data in joint-probe analysis and forecasting. It is written in Python and compresses the $\ell<30$ temperature and polarization angular power spectra information from {\it Planck} into two log-normal bins in temperature and three in polarization. 
These angular scales constrain the optical depth to reionization and provide a lever arm to constrain the tilt of the primordial power spectrum. 
We show that cosmological constraints on $\Lambda$CDM model parameters using \texttt{Planck-low-py} are consistent with those derived with the full \texttt{Commander} and \texttt{SimAll} likelihoods from the {\it Planck} legacy release.
\end{abstract}

\maketitle

\section{Introduction}

The current best fit model of the universe is \LCDM, which provides a remarkable fit to a variety of cosmological data with only six cosmological parameters. The cosmic microwave background (CMB) radiation from the early universe provides strong constraints on these parameters, and the CMB temperature and polarization have been measured over the full sky by the \textit{Planck} satellite, with its legacy data released in 2018 \citep{planck2018_like,planck2018_cosmo}. Various current and planned CMB experiments are focused on measuring the CMB from the ground, improving on \textit{Planck} in resolution and  
instrumental noise.
However, because of atmospheric noise and limited sky coverage, it is challenging to measure large angular scale CMB fluctuations from the ground. 
Both galaxy surveys and ground-based CMB data will thus continue to benefit in the coming decade from including the \textit{Planck} data in constraints, especially on large scales.

The \textit{Planck} legacy release includes likelihood functions describing the temperature and $E$-mode polarization data on large scales, where the probability distribution of the angular power spectrum is non-Gaussian \citep{planck2018_like}. While the evaluation time of these likelihoods for the {\textit{Planck}} low-$\ell$ data is not a limiting factor compared to the time needed, for example, to run Boltzmann codes, the existing package includes code in both C and Fortran 90. 
 To facilitate the ease of use of the data, for both analysis and forecasting applications, we thus present an alternative option that compresses the data to a set of independent log-normal bins. The low-$\ell$ (large scale) $E$-mode polarization primarily constrains the optical depth to reionization, $\tau$, and the low-$\ell$ temperature provides a lever arm to constrain the tilt of the primordial power spectrum, $n_s$. For most models of interest, one therefore 
only needs a few numbers to describe the \textit{Planck} low-$\ell$ data
\citep[e.g.,][]{heavens_massive_2000, alsing_generalized_2018}. 

This log-normal compression extends work presented in \citet{Prince2019} which included a compression of the temperature spectrum. Here we use two bins to describe the low-$\ell$ temperature power spectrum 
and three 
bins for $E$-mode polarization, 
leading to a simple likelihood code written in Python that is portable and easy to combine with other cosmological data. The method could also be applied to the alternative {\textit{Planck}} data processing from the \texttt{NPIPE} maps described in \citet{planck_npipe:2020}.
Our code, \texttt{Planck-low-py}, is publicly available on Github\footnote{\url{https://github.com/heatherprince/planck-low-py}}. It can be used in combination with \texttt{Planck-lite-py}\footnote{\url{https://github.com/heatherprince/planck-lite-py}}, our Python implementation of the \textit{Planck} team's \texttt{Plik\_lite} foreground-marginalized likelihood for the high-$\ell$ data. Use of either code should reference the \citet{planck2018_like} data. 
Other Python implementations of \textit{Planck} likelihoods, including \texttt{Plik\_lite}, are also already available with the public cosmological sampling code \texttt {cobaya}\footnote{\url{https://cobaya.readthedocs.io/en/latest/likelihood_planck.html}}. 

We describe the \textit{Planck} data and our low-$\ell$ binned likelihoods in section \ref{sec:planck}. In section \ref{sec:params} we compare \LCDM\ parameter constraints from \texttt{Planck-low-py} with those from the legacy \textit{Planck} likelihoods; extended models were discussed in \cite{Prince2019}. We conclude in section \ref{sec:conclude}.

\section{The \textit{Planck} likelihood and low-$\ell$ binning}
\label{sec:planck}

\begin{figure*}[!thbp]
  \centering
  \includegraphics[width=\linewidth]{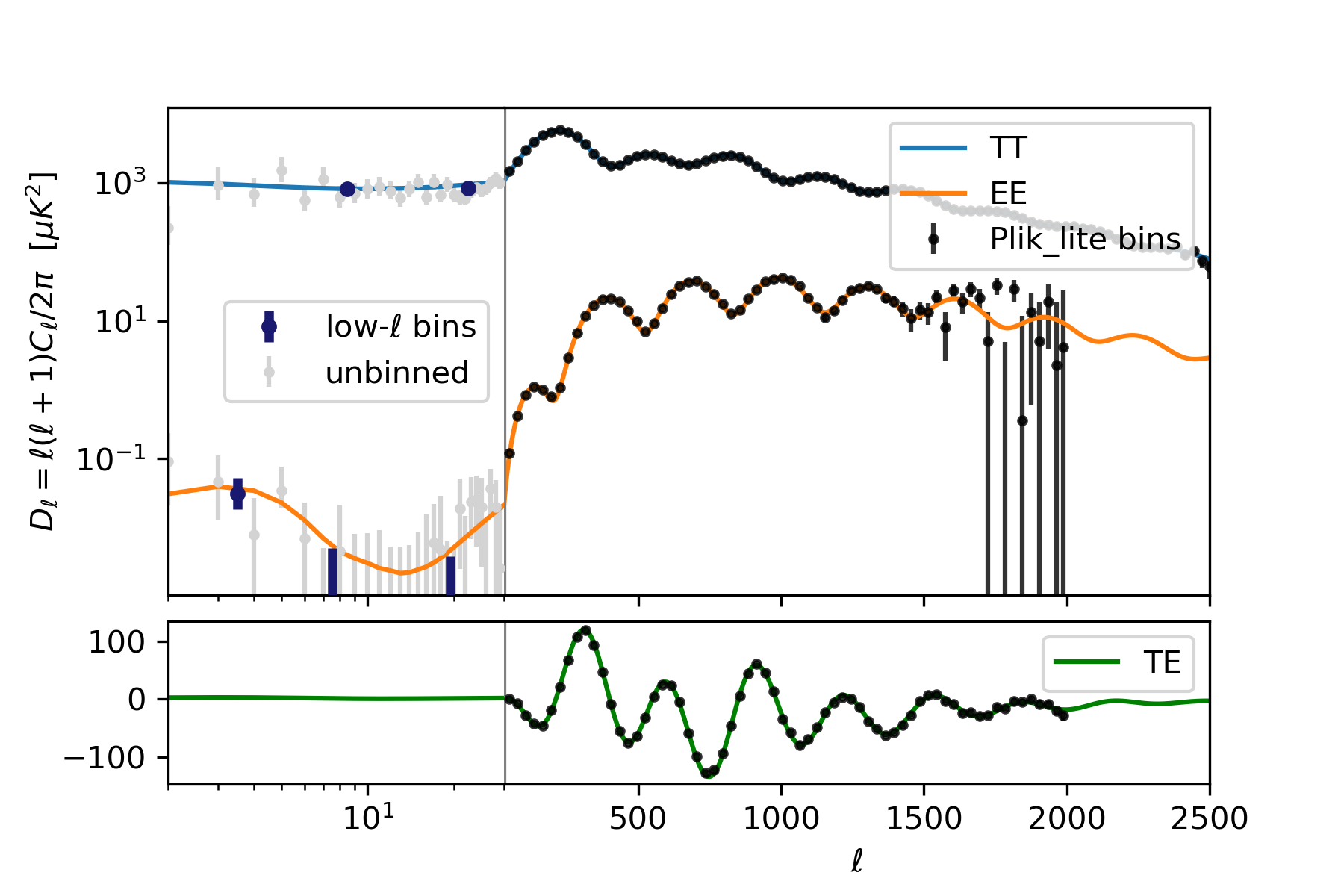}
  \caption{The \textit{Planck} 2018 temperature and polarization power spectra and best-fit theory curves, from \citet{planck2018_like}. The log-normal binned $\ell<30$ $TT$ and $EE$ spectra estimated in this paper, described in section \ref{sec:planck}, are shown on the left of the upper panel. They compress the per-$\ell$ likelihoods (grey markers) described by the {\it Planck} \texttt{Commander} and \texttt{SimAll} likelihoods respectively. The Gaussian, foreground-marginalized $\ell \geq 30$ data (black markers on the right) are used in the 
  \texttt{Plik\_lite} cosmological likelihood and are shown for reference. The low-$\ell$ $TE$ cross spectrum is not used in the {\it Planck} cosmological analysis.}
   \label{fig:spectra}
\end{figure*}

The  likelihood function used for \textit{Planck}'s cosmological analysis, describing the probability of the data given some model, is separated into two regimes, with different approaches for large and small angular scales \citep{planck2018_like}.

At $\ell \geq 30$ (corresponding to scales smaller than several degrees on the
sky) the likelihood $\mathcal{L}$ for the temperature and $E$-mode polarization
power spectra and cross spectrum ($TT$, $EE$ and $TE$) is modeled as a Gaussian distribution, with
\be
-2 \ln \mathcal{L} = (C_b^{\rm th} - C_b^{\rm data})^T Q^{-1}(C_b^{\rm th} -
C_b^{\rm data}),
\ee
to within an overall additive constant, with binned data $C_b^{\rm data}$,
binned theory $C_b^{\rm th}$, and binned covariance matrix $Q$. For the
\texttt{Plik\_lite} likelihood \citep{planck2018_like}, these data spectra represent an
estimate of the CMB bandpowers, with foregrounds already marginalized over.
The bandpowers for these binned  high-$\ell$ data are shown in Fig.~\ref{fig:spectra}, as the black markers in the right panels.
The multi-frequency high-$\ell$ \texttt{Plik} likelihood does not pre-marginalize over foregrounds, and gives results in agreement with \texttt{Plik\_lite}.

At $\ell<30$ (large angular scales on the sky) the distribution of the angular power spectrum is non-Gaussian \citep[see, e.g.,][]{bond/jaffe/knox:2000}. The {\it Planck} low-$\ell$ temperature likelihood is derived using the \texttt{Commander} framework \citep{eriksen/etal:2008}, which uses Gibbs sampling to explore the joint distribution of the CMB temperature map, CMB temperature angular power spectrum, and foreground parameters, and then uses a Gaussianized Blackwell-Rau estimator to describe the likelihood of the modeled temperature angular power spectrum given the data.

The {\it Planck} low-$\ell$
polarization uses a separate \texttt{SimAll} likelihood built from simulations \citep{planck2018_like}. This likelihood uses the $E$-mode polarization angular cross spectrum between the $100$~GHz and $143$~GHz channels, computed using the quadratic maximum likelihood (QML) approach \citep{tegmark/etal:2001,efstathiou:2006}, with templates for synchrotron and dust contamination used to remove the foregrounds. 
The likelihood is then constructed from this data power spectrum using a suite of simulations.

The curl-like polarization $B$-modes are not used for the baseline {\it Planck} cosmological analysis. The large scale temperature-polarization correlation (captured in the low-$\ell$ $TE$ cross spectrum) is also excluded due to performing poorly in null tests, indicating that some systematic or foreground effects remain unaccounted for.
In this paper we thus focus just on compressing the $\ell<30$ temperature and $E$-mode polarization likelihoods.

\subsection{Low-$\ell$ temperature bins}

\begin{figure}[!th]
  \centering
  \includegraphics[width=\linewidth]{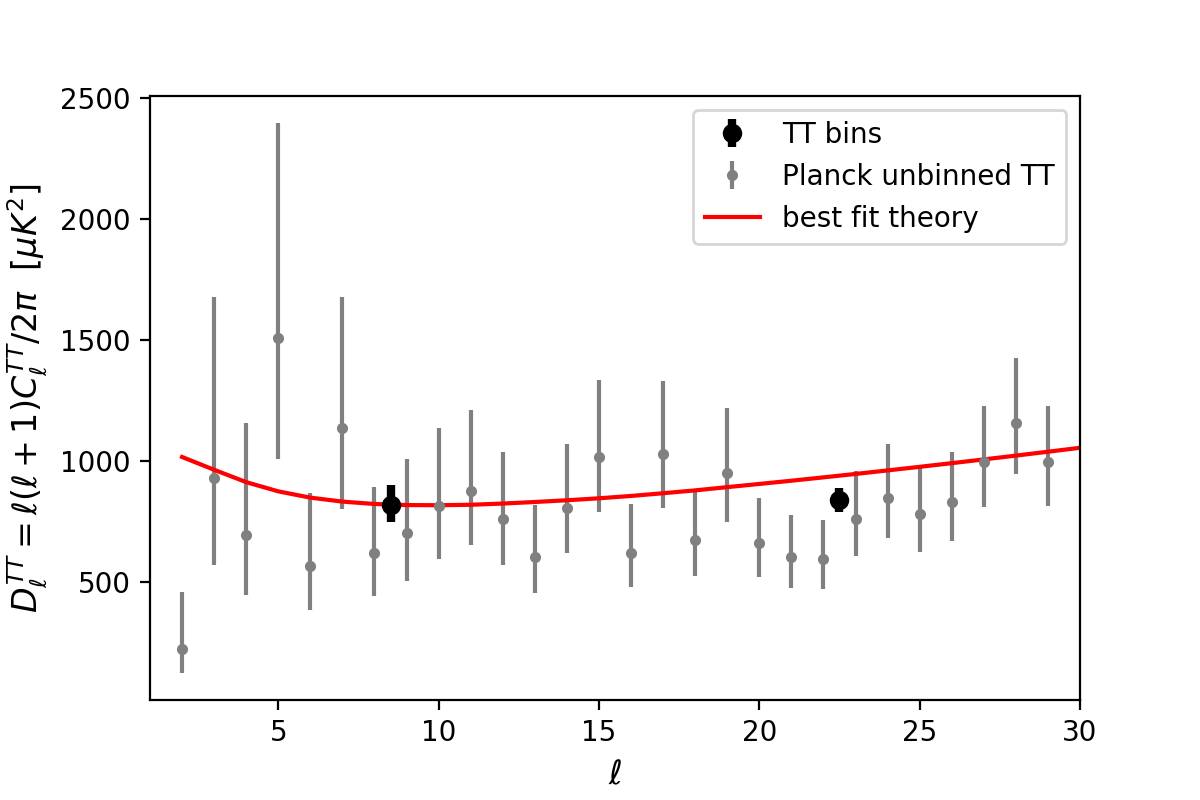}
  \caption{The low-$\ell$ \textit{Planck} temperature power spectrum $D_\ell^{TT}=\ell(\ell+1) C_\ell^{TT} /2\pi$. The binned power and errors are shown in black. The \textit{Planck} unbinned power spectrum and errors are in grey. The errorbars are asymmetric because on these scales the power spectrum is non-Gaussian. The value of the power spectrum is taken from the peak of the probability distribution and the errorbars come from the half-maximum values. The theory curve for the best fit $\Lambda$CDM model is shown in red. The power spectrum is close to flat on these large angular scales due to the Sachs-Wolfe effect.}
   \label{fig:low_tt_spec}
\end{figure}

The low-$\ell$ temperature power spectrum, shown in Fig.~\ref{fig:low_tt_spec}, is approximately flat due to the Sachs-Wolfe effect
\citep{sachs/wolfe:1967}. 
In the $\Lambda$CDM model, which is an excellent fit to \textit{Planck} data, the low-$\ell$ temperature power provides a large-scale anchor for constraining the spectral index of the primordial fluctuation power spectrum, $n_s$, which tilts the resulting temperature power spectrum. 
Within simple models, including $\Lambda$CDM, the lack of complicated structure on these large scales suggests that we could bin the low-$\ell$ data without losing  
information about the cosmological parameters. In \citet{Prince2019} we demonstrated that compressing to two bins was sufficient to reproduce parameters for the {\it Planck} 2015 temperature data. There the compresssion to two Gaussian bins 
was useful for the data compression technique applied to the full $\ell$ range of the {\it Planck} power spectrum.

\begin{figure}[!th]
  \centering
  \includegraphics[width=\linewidth]{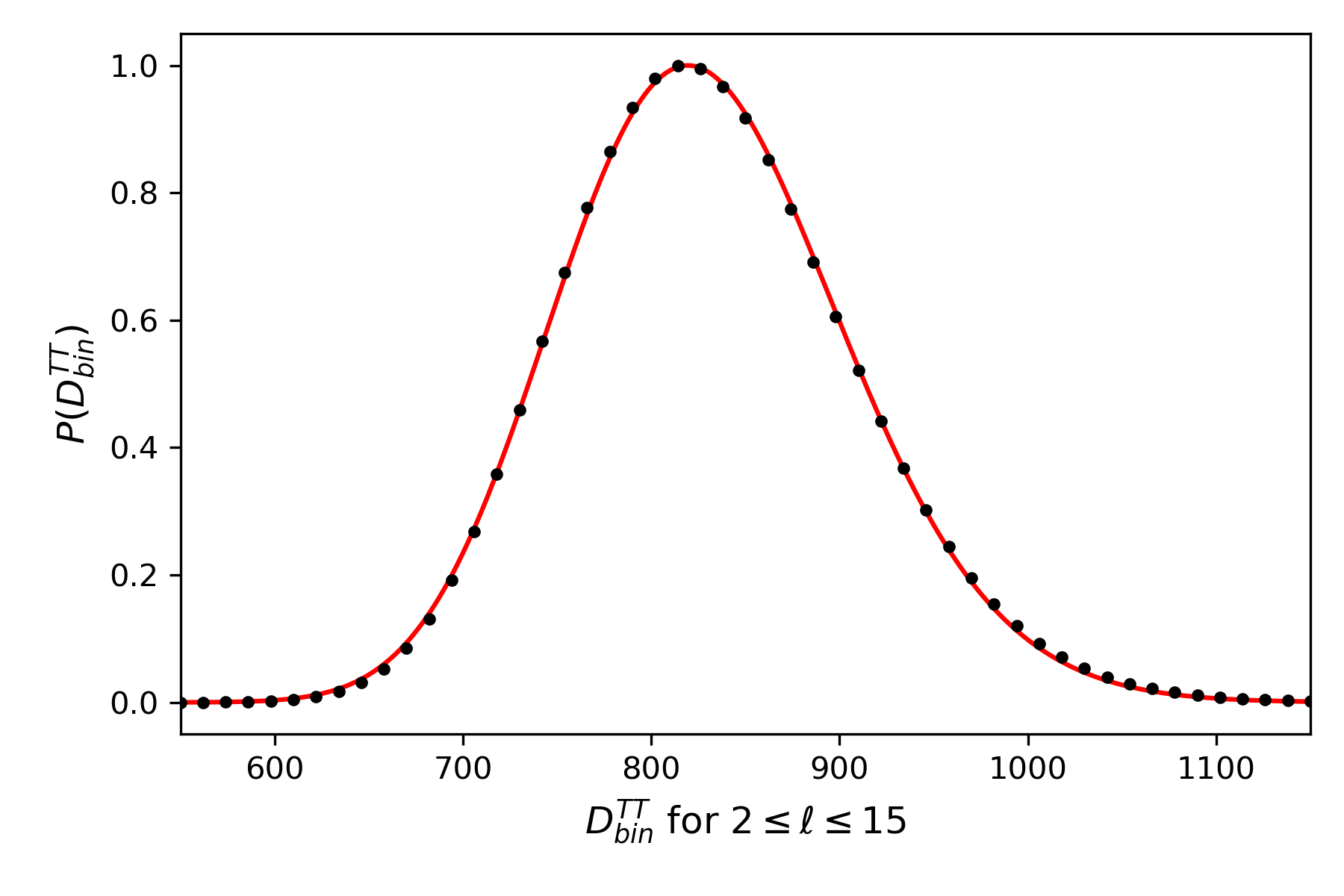}
  \includegraphics[width=\linewidth]{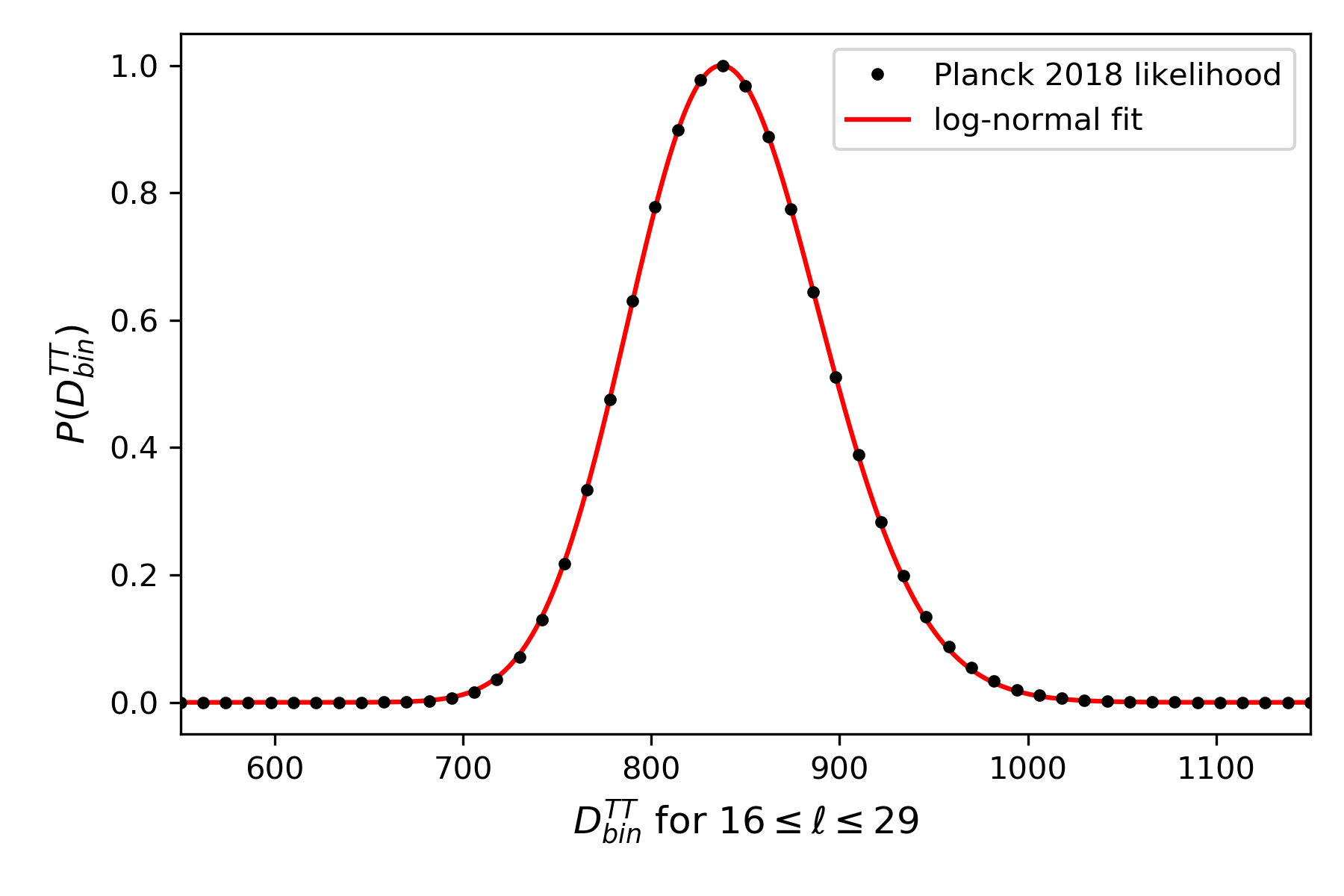}
  \caption{The posterior distribution for $D_\ell^{TT}=\ell(\ell+1) C_\ell^{TT} /2\pi$ for two low-$\ell$
  temperature bins ($\ell=$ 2-15 and 16-29) using the \textit{Planck} 2018 Commander low-$\ell$ temperature
  likelihood. The best-fitting log-normal distributions are shown in red.}
   \label{fig:lowell}
\end{figure}

Here we refine this compression using a log-normal distribution to approximate the likelihood, motivated by e.g., \cite{bond/jaffe/knox:2000}.  We find that the 
log-normal bins give a slightly better fit to the shape of the posterior of the power in each bin, 
and is also consistent with how we treat the low-$\ell$ polarization data in Section~\ref{sec:lowEE}.

As in \citet{Prince2019} we form two temperature bins, 
one for $2 \leq \ell \leq 15$ and one for $16 \leq \ell \leq 29$. We find the probability distribution for the power spectrum $D_\ell = \ell (\ell+1) C_\ell /2 \pi$ by conditionally sampling the posterior distribution for the power in each bin using the \textit{Planck} 2018 \texttt {Commander} likelihood using the \texttt{Cobaya} cosmological sampling code \citep{cobaya2020}, estimating 
\be
p(\theta|d) \propto p(d|\theta)p(\theta).
\ee
Here the parameters $\theta$ are the binned values $D_{2\leq \ell\leq 15}$ and
$D_{16\leq \ell\leq 29}$,
assuming a constant value for $D$ in each bin. We assume
uniform priors on $\theta$.

The probability distributions for the two low-$\ell$ power spectrum bins are shown in Fig. \ref{fig:lowell}, together with the best-fit log-normal probability distributions. A log-normal distribution for the likelihood of $D_{\rm bin}$ means that $\ln(D_{\rm bin}$) is close to normally distributed, or
\be
\mathcal{L}(x) = p(x)= \frac{1}{x \sigma \sqrt{2\pi}} e^{-(\ln x-\mu)^2/(2\sigma^2)},
\label{eq:lognorm}
\ee
for $x=D_{\rm bin}$. Although there is some motivation for using a combination of log-normal and Gaussian distributions to describe the probability distribution of the unbinned angular power spectra \citep{wmap2003}, we find that the log-normal fit to the bins gives an acceptable fit and accurately reproduces parameter constraints. The best-fitting parameters 
for the two  
bins are 
\ba
\mu_1=6.717, ~~~ \sigma_1=0.09247, \nonumber \\
\mu_2=6.734, ~~~ \sigma_2=0.06038.
\ea

The covariance between bins is small, so we treat the likelihood of each bin independently. 
To compute the likelihood for a given theory power spectrum, we convert the theory $C^{\rm th}_\ell$ to $D^{\rm th}_\ell$, bin $D^{\rm th}_\ell$ into two bins, $D_1^{TT}$ and $D_2^{TT}$, compute the log-normal likelihood for each bin, and then multiply them together (corresponding to adding the log-likelihoods), giving 
\begin{align}
\ln \mathcal{L}^{TT} =& \ln \mathcal{L}_1^{TT} + \ln \mathcal{L}_2^{TT} \nonumber \\
=& \ln \left( \frac{1}{D_1^{TT} \sigma_1 \sqrt{2\pi}} e^{-(\ln D_1^{TT}-\mu_1)^2/(2\sigma_1^2)} \right)~~ + \nonumber\\
& \ln \left(\frac{1}{D_2^{TT} \sigma_2 \sqrt{2\pi}} e^{-(\ln D_2^{TT}-\mu_2)^2/(2\sigma_2^2)} \right).
\end{align}
This compression approximates the cosmic variance contribution to the uncertainty on the power spectrum as being independent of the theory, as in the $\ell>30$ {\it Planck} likelihood.

The bandpowers for these two log-normal bins are indicated in
Fig. \ref{fig:low_tt_spec} (the large black points), together with the
unbinned low-$\ell$ power spectrum.
For plotting purposes 
the $D_{\rm bin}$
at the peak of the distribution (the mode) is shown, which for a log-normal distribution is at
$D_{\rm bin}=e^{\mu-\sigma^2}$.
We plot errors using the $D$ values at which the probability 
drops to $0.61$ of its maximum, which would be at $1\sigma$ for a Gaussian distribution, showing
\ba
D_{2\leq \ell\leq 15}^{TT}=819^{+79}_{-72} ~ \mu{\rm K}^2, \nonumber \\
D_{16\leq \ell\leq 29}^{TT}=837^{+52}_{-49} ~ \mu{\rm K}^2.
\ea

\subsection{Low-$\ell$ $E$-mode polarization}
\label{sec:lowEE}

\begin{figure}[!th]
  \centering
  \includegraphics[width=\linewidth]{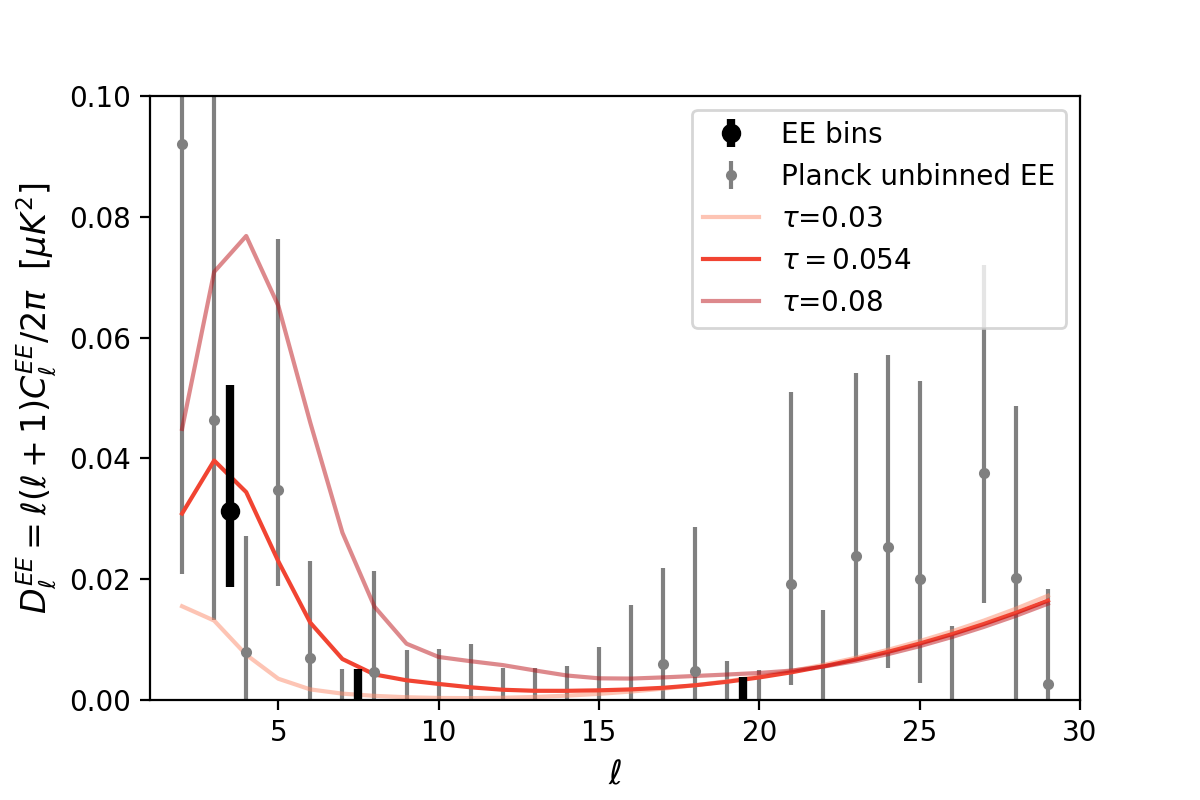}
  \caption{The low-$\ell$ E-mode polarization power spectrum, with the $D^{EE}$ estimated in three bins ($\ell=$ 2-5, 6-9 and 10-29) shown in black. The \textit{Planck} per-$ell$ power spectrum is shown in grey. The values and errors for these non-Gaussian data points come from the peak and half-maximum of the conditional posterior distribution for each $D_\ell$.  The theory curve for the best fit model (red) and for high and low values of the optical depth to reionization $\tau$ are shown. The reionization bump in the power spectrum comes from polarization caused by Thompson scattering off free electrons since reionization. This signal is higher for greater $\tau$.}
   \label{fig:low_ee_spec}
\end{figure}

\begin{figure}[!thb]
  \centering
  \includegraphics[width=\linewidth]{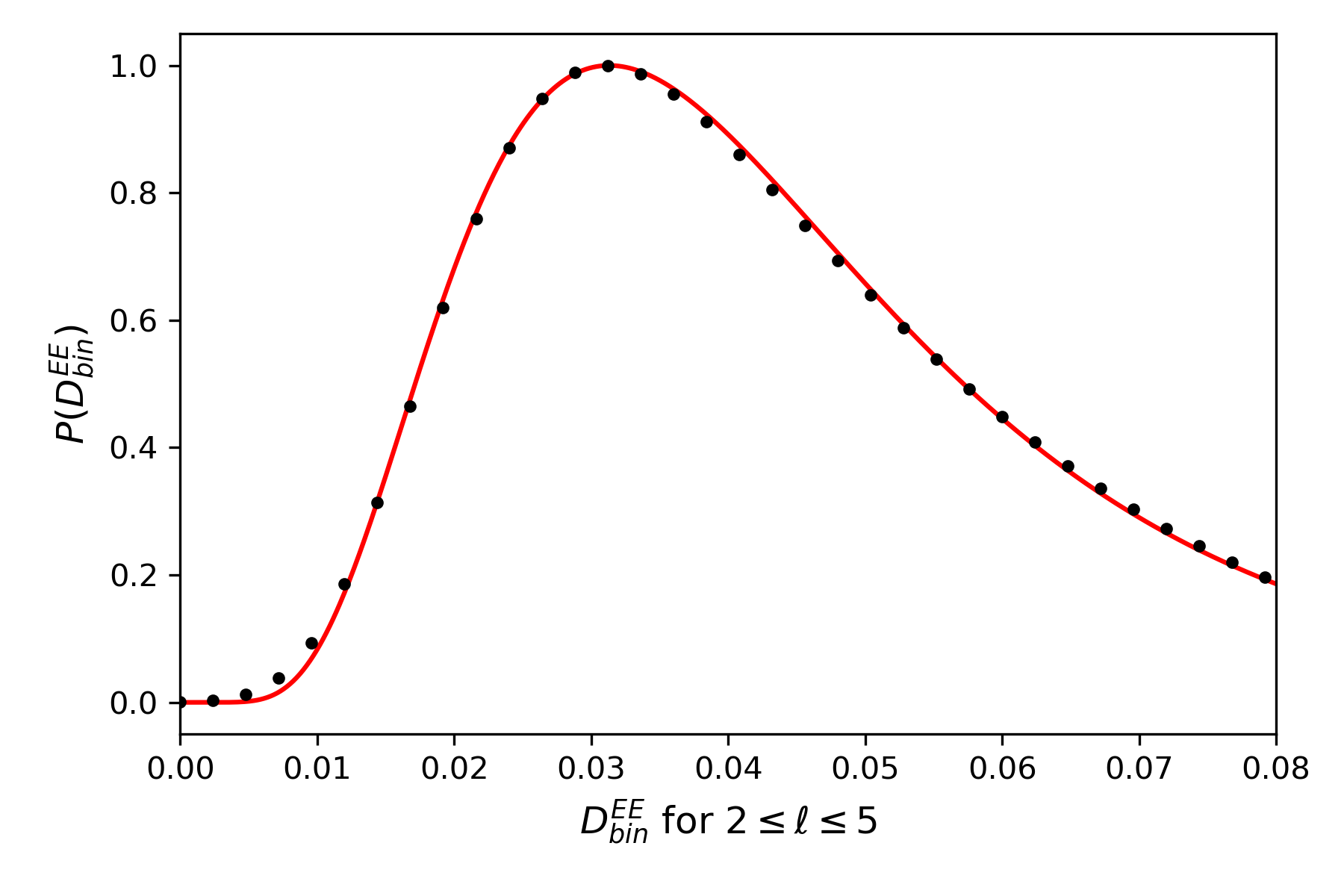}
  \includegraphics[width=\linewidth]{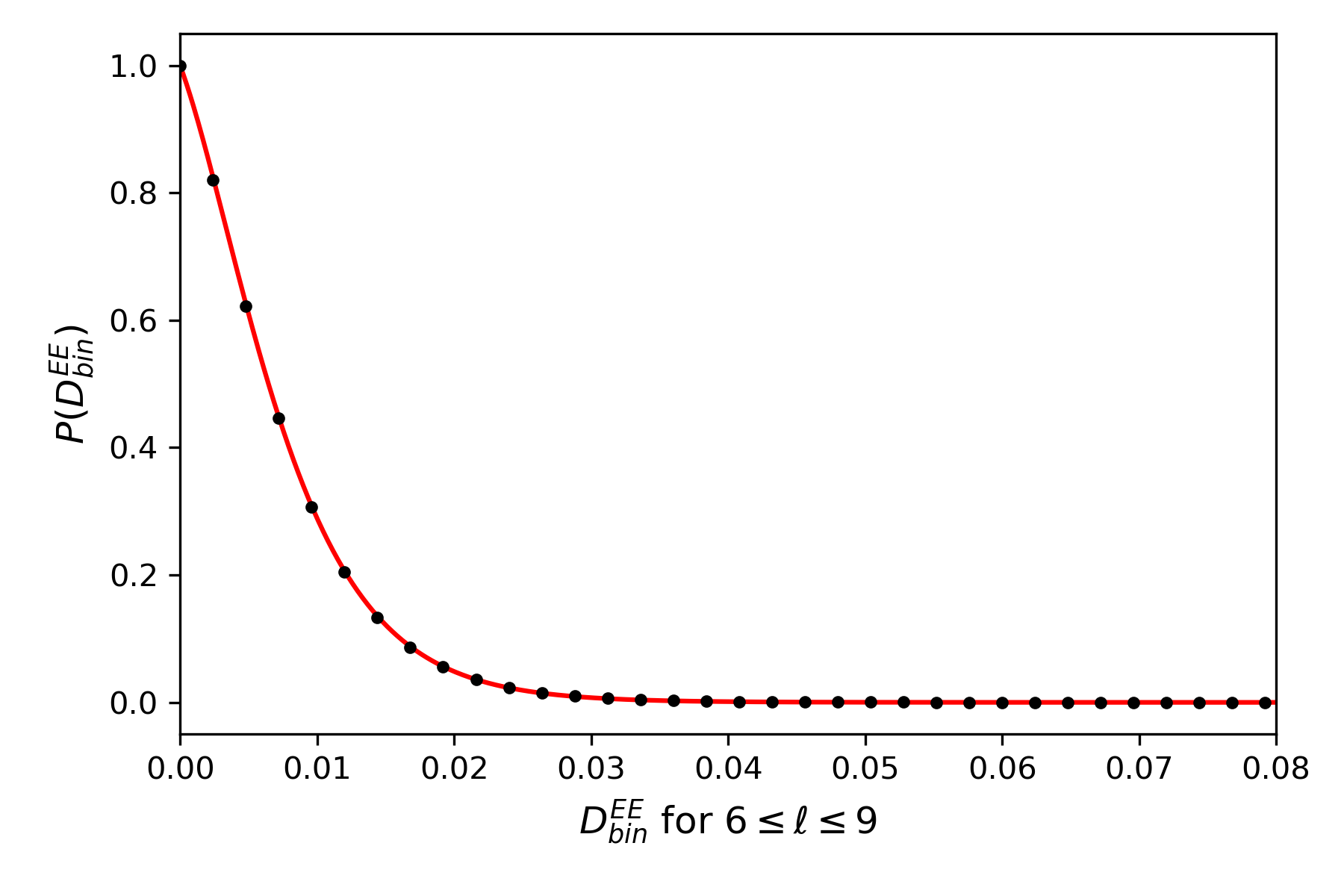}
  \includegraphics[width=\linewidth]{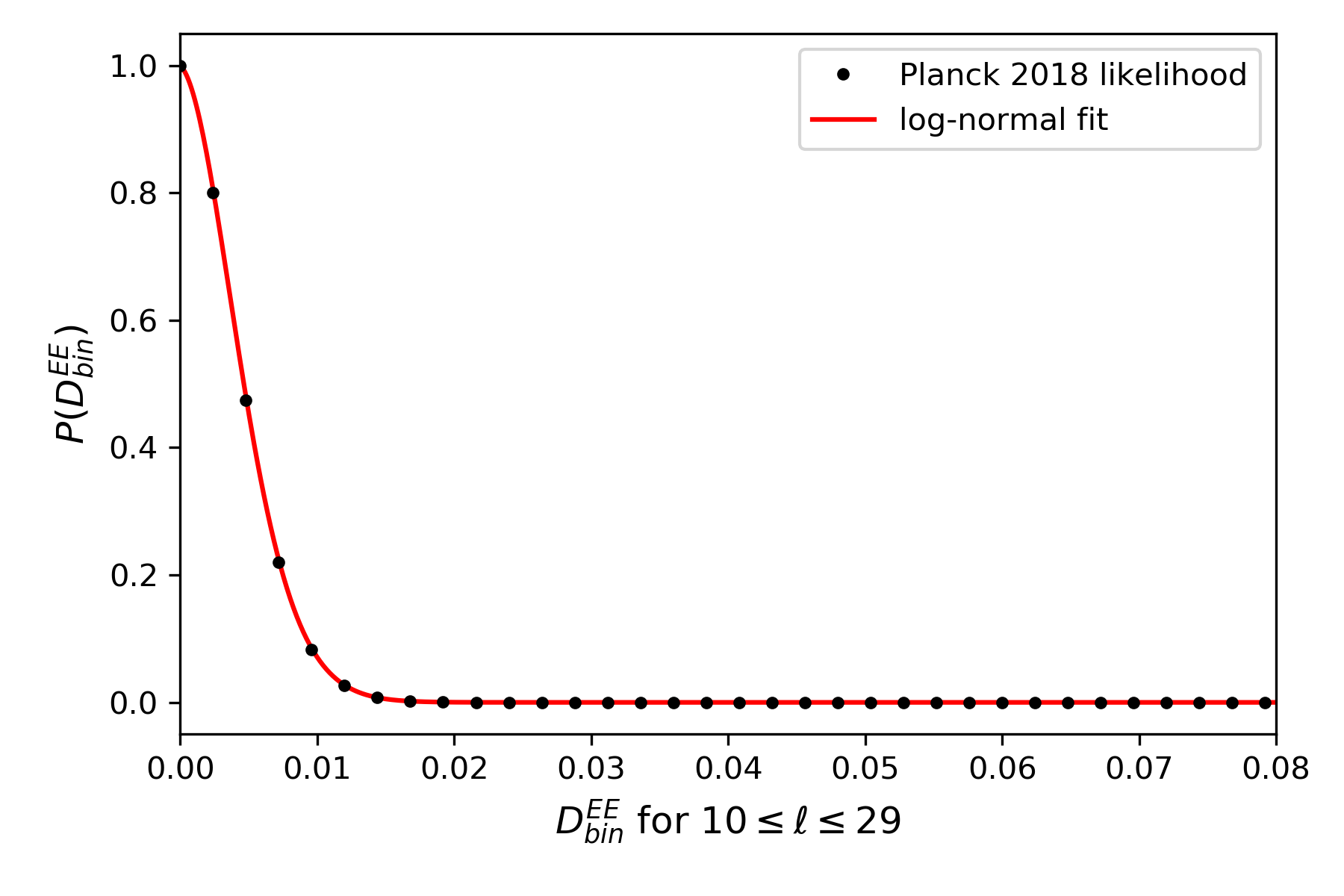}
  \caption{The conditional posterior distribution for $D_\ell^{EE}=\ell(\ell+1) C_\ell^{EE} /2\pi$ for three low-$\ell$
  E-mode polarization bins, estimated using the \textit{Planck} 2018 \texttt{SimAll} low-$\ell$ polarization  likelihood. The best-fit offset log-normal distribution is shown in red for each bin. The $\ell$ ranges of the three bins ($\ell=$ 2-5, 6-9 and 10-29) are chosen to capture the low-$\ell$ polarization features imprinted from varying the optical depth to reionization, $\tau$.}
   \label{fig:lowell_ee_bins}
\end{figure}

The CMB $E$-mode polarization signal is created by Thomson scattering of CMB photons off electrons, both at the last scattering surface and during reionization when electrons are once again free \citep{bond/efstathiou:1984,zaldarriaga:1997}. Thomson scattering during reionization creates large-scale $E$-mode polarization, the amplitude of which depends primarily on the
optical depth to reionization, $\tau$. This `reionization bump' can be seen in Fig.~\ref{fig:low_ee_spec}, which shows the $\Lambda$CDM theoretical $EE$ power spectrum for a few different values of $\tau$, including the best-fit $\tau=0.054$ from \cite{planck2018_cosmo}. As the optical depth to reionization increases, so does the low-$\ell$ EE power. 

The \textit{Planck} 2018 low-$\ell$ polarization likelihood was built using simulations to determine the probability distributions for the $EE$ and $BB$ spectra. 
Here we compress the low-$\ell$ $EE$-mode polarization data into three bins, one that primarily constrains the height of the reionization bump  using $2\leq \ell \leq 5$, one that constrains the width of the bump using $6 \leq \ell \leq 9$, and one wider bin for $10 \leq \ell \leq 29$ where the $EE$ power spectrum has less structure. We do not compress the $B$-mode polarization data, as it is not used in the main {\it Planck} likelihood combination for cosmology constraints.

The probability distribution for the power in each bin is shown in Fig.~\ref{fig:lowell_ee_bins}. We find that the distributions are well described by an offset log-normal distribution, which modifies equation (\ref{eq:lognorm}) for the probability distribution of $x=D_{\rm bin}^{EE}$ to
\be
\mathcal{L}(x) = p(x)= \frac{1}{(x-x_0) \sigma \sqrt{2\pi}} e^{-(\ln (x-x_0) -\mu)^2/(2\sigma^2)}.
\label{eq:lognorm_shifted}
\ee
 The sharp drop off from $D=0$, giving an upper limit, for the second two of the three bins is better described by this offset log-normal distribution, although for the first bin an unshifted log-normal ($x_0=0$) fits well. The best-fit parameters for the log-normal bins 
 are 
\ba
&\mu_1^{EE}=-3.202, ~~~ \sigma_1^{EE}=0.5114, ~~~x_{0,1}^{EE}=0\nonumber \\
&\mu_2^{EE}=-3.997, ~~~ \sigma_2^{EE}=0.3458, ~~~x_{0,2}^{EE}=-0.01851 \nonumber \\
&\mu_3^{EE}=-3.358, ~~~ \sigma_3^{EE}=0.1138, ~~~x_{0,3}^{EE}=-0.03462.
\ea

\begin{figure*}[!bth]
  \centering
  \includegraphics[width=\linewidth]{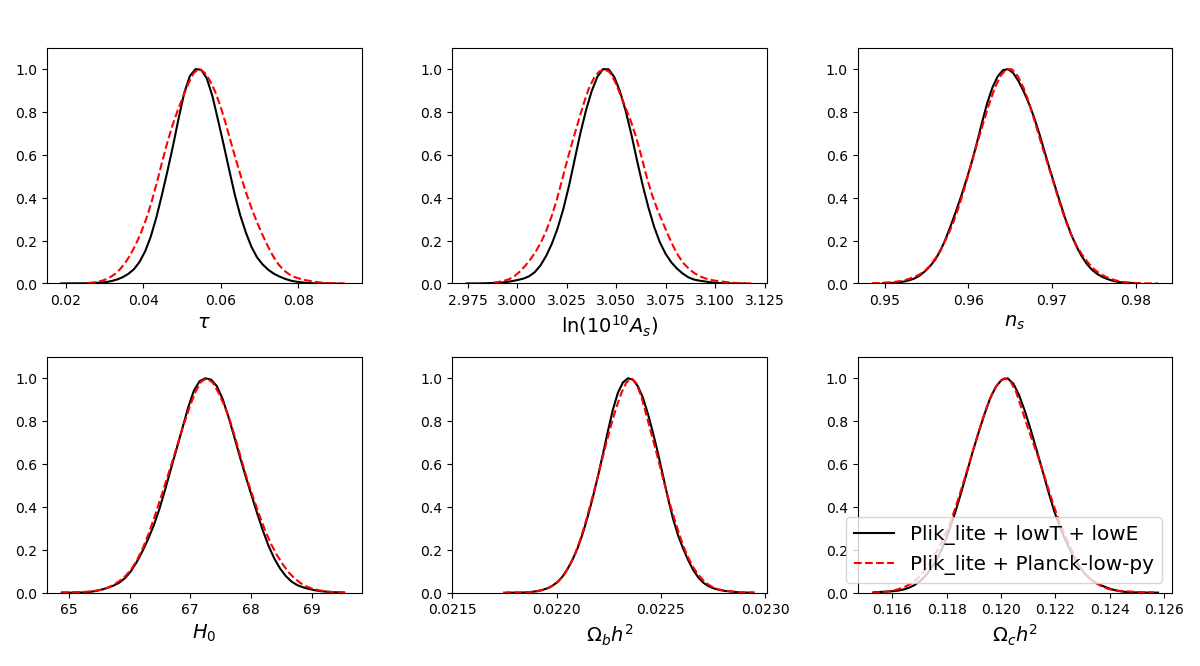}
  \caption{Posterior distributions of the $\Lambda$CDM parameters estimated using the
  low-$\ell$ \textit{Planck} 2018 temperature (\texttt{Commander}) and EE (\texttt{SimAll}) likelihoods 
  (black), compared to those using \texttt{Planck-low-py} (our log-normal compressed low-ell likelihood) (red dashed). In both cases \texttt{Plik\_lite} is used at $\ell>30$.  The parameter constraints agree to within 0.1$\sigma$.
  The constraints on $\tau$ and $A_s$ from \texttt{Planck-low-py} are slightly broader than the \texttt{Commander + SimAll} constraints, likely due to some loss of information from compressing the data.}
   \label{fig:params_two_low_ell_bins}
\end{figure*}

As for the temperature, we neglect the covariance between bins, so the log-normal likelihoods from the three $EE$ bins are combined independently to give the compressed low-$\ell$ $E$-mode polarization likelihood:
\be
\ln \mathcal{L}^{EE} = \ln \mathcal{L}_1^{EE} + \ln \mathcal{L}_2^{EE}  + \ln \mathcal{L}_3^{EE} .
\ee

These best-fitting parameters are used to plot the mode and errors for the first bin in Fig.~\ref{fig:low_ee_spec}, with
\be
D_{2\leq \ell\leq 5}^{EE}  = 0.031^{+0.021}_{-0.013} ~ \mu {\rm K}^2,
\ee
and the upper limits, where the probability drops to $0.61$ of maximum, for the second and third bins shown at
\ba
D_{6\leq \ell\leq 9}^{EE}  \leq 0.005 ~ \mu {\rm K}^2, \nonumber \\
D_{10\leq \ell\leq 29}^{EE} \leq 0.004 ~ \mu {\rm K}^2.
\ea

\subsection{Software products}

We release a public Python likelihood code, \texttt{Planck-low-py},\footnote{\url{https://github.com/heatherprince/planck-low-py}} which uses these independent log-normal likelihoods to describe five \textit{Planck} low-$\ell$ temperature and polarization bins. \texttt{Planck-low-py} has 12 parameters: two for each of the two $TT$ bins and eight in total for the $E$-mode polarization (two for the first log-normal bin and three each for the other two offset log-normal bins). The temperature and polarization likelihoods can be used separately or together. 

This likelihood code provides a simple 
alternative to the full \textit{Planck} 
low-$\ell$ likelihood functions, for models without unusually complex large-scale behavior. Any use of it should reference the \citet{planck2018_like} data.
The code can be used in combination with any $\ell>30$ {\it Planck} likelihoods, including
our Python implementation of \texttt{Plik\_lite}, \texttt{planck-lite-py}\footnote{\url{https://github.com/heatherprince/planck-lite-py}}, or the Python implementations of \textit{Planck}'s likelihoods in \texttt{Cobaya}.

\section{Parameter constraints}
\label{sec:params}

We test the compressed \texttt{Planck-low-py} by comparing the constraints on $\Lambda$CDM parameters with those from the full \textit{Planck} 2018 low-$\ell$ $TT$ and $EE$ likelihoods, using the same high-$\ell$ \texttt{Plik\_lite} likelihood in both cases.
The results are plotted in Fig. \ref{fig:params_two_low_ell_bins}, which shows the
posterior probabilities for the six $\Lambda$CDM  parameters (the Hubble constant,
baryon density, cold dark matter density, amplitude and spectral index of
primordial fluctuations, and optical depth to reionization). 
These parameter constraints were obtained using \texttt{Cobaya} \citep{cobaya2020}\footnote{\url{https://ascl.net/1910.019}} with the {\texttt CAMB} cosmological Boltzmann code \citep{Lewis:1999bs,Howlett:2012mh} and the MCMC sampler developed for \texttt{CosmoMC} \citep{Lewis:2002ah,Lewis:2013hha}
(with the `fast dragging' procedure described by \citet{Neal:2005}). 

The reference constraints (black) are obtained from the high-$\ell$ temperature and polarization foreground-marginalized \texttt{Plik\_lite} 
likelihood, the low-$\ell$ EE likelihood (\texttt{SimAll}), and the low-$\ell$ temperature-only likelihood (\texttt{Commander}) of \textit{Planck}'s 2018 data release \citep{planck2018_like}. The red dashed lines show our constraints from combining \texttt{Plik\_lite} with our \texttt{Planck-low-py} compressed likelihood.

We find that all of the $\Lambda$CDM cosmological parameters are consistent to within 0.1$\sigma$. The constraints on the optical depth to reionization $\tau$ from Planck-low-py is slightly broader than, but consistent with, the \textit{Planck} 2018 results from the full low-$\ell$ temperature and polarization likelihoods (\texttt{Commander} and \texttt{SimAll}) The amplitude of primordial fluctuations $A_s$ is correlated with $\tau$, so its constraints are also broadened. The constraints on the other $\Lambda$CDM parameters are equivalent to the \textit{Planck} constraints.
We showed in \cite{Prince2019} that a compression of the low-$\ell$ temperature is also effective for a simple extension to $\Lambda$CDM.

We find that using just one or two lognormal bins to describe the low-$\ell$ $EE$ likelihood is insufficient to reproduce the parameter constraints, with three bins better  
capturing the structure of the reionization bump (see Fig.~\ref{fig:low_ee_spec}).
The \texttt{Planck-low-py} $E$-mode likelihood can also be used as an alternative to a $\tau$ prior, with the advantage of directly using the amplitude of the EE power spectrum.
For the equivalent of imposing a broader prior on $\tau$, as done in \cite{aiola/etal:2020} for example, one could inflate the error bars of the $EE$ bins.

\section{Discussion}
\label{sec:conclude}

We have demonstrated that the \textit{Planck} 2018 low-$\ell$ temperature and $E$-mode polarization data can be effectively compressed to two and three log-normal bins respectively while retaining accurate constraints on $\Lambda$CDM cosmological parameters. This compression would also be appropriate for non-$\Lambda$CDM models which do not have unusual large-scale features. We present a public Python likelihood code, \texttt{Planck-low-py}, which uses a total of 12 numbers (4 for temperature, 8 for polarization) to represent the \textit{Planck} low-$\ell$ data and which can be used as a light-weight alternative to the \texttt{Commander} and \texttt{SimAll} likelihoods. 

The \textit{Planck} data will provide the community with the tightest constraint on the optical depth to reionization $\tau$, and the leading large-scale temperature measurements,  for some time. 
Light-weight versions of the \textit{Planck} low-$\ell$ likelihoods can thus be useful for future explorations of combined datasets, as well as for forecasting purposes in the design of upcoming experiments.

\section{Acknowledgments}
We thank Erminia Calabrese and Adri Duivenvoorden for useful suggestions. We acknowledge the use of 
data and code from the Planck Legacy Archive. JD gratefully acknowledges support from the Institute for Advanced Study. 

\input{main.bbl}

\end{document}

%% file: main.bbl
%